# Channel Impulse Response Peak Clustering Using Neural Networks

Petr Horky[1], Ales Prokes[1], Radek Zavorka[1], Josef Vychodil[1], Jan Marcin Kelner[2], Cezary Henryk Ziołkowski[2], Aniruddha Chandra[3]

[1] Department of Radio Electronics, Brno University of Technology, Brno, Czech Republic
[2] Institute of Communications Systems, Faculty of Electronics, Military University of Technology, 00-908 Warsaw, Poland
[3] National Institute of Technology, Durgapur, India
email: 244016@vutbr.cz

*Abstract*—This paper introduces an approach to process channel sounder data acquired from Channel Impulse Response (CIR) of 60GHz and 80GHz channel sounder systems, through the integration of Long Short-Term Memory (LSTM) Neural Network (NN) and Fully Connected Neural Network (FCNN). The primary goal is to enhance and automate cluster detection within peaks from noised CIR data. The study initially compares the performance of LSTM NN and FCNN across different input sequence lengths. Notably, LSTM surpasses FCNN due to its incorporation of memory cells, which prove beneficial for handling longer series.

Additionally, the paper investigates the robustness of LSTM NN through various architectural configurations. The findings suggest that robust neural networks tend to closely mimic the input function, whereas smaller neural networks are better at generalizing trends in time series data, which is desirable for anomaly detection, where function peaks are regarded as anomalies.

Finally, the selected LSTM NN is compared with traditional signal filters, including Butterworth, Savitzky-Golay, Bessel/Thomson, and median filters. Visual observations indicate that the most effective methods for peak detection within channel impulse response data are either the LSTM NN or median filter, as they yield similar results.

*Keywords*— LSTM, FCNN, DBSCAN, anomaly detection, clusters, peak detection, channel impulse response

## I. INTRODUCTION

In recent times, the concept of Artificial Intelligence (AI) has gained significant prominence, with continual exploration of novel domains for its application and utilization. AI is a very broad term and one of the branches of AI is Machine Learning (ML). Using machine learning algorithms can lead to very convenient results and it is a useful tool to be used in algorithms because the process of evaluating data can be described mathematically. A specific group of ML algorithms is Neural Networks, which if they reach a certain complexity are called Deep Learning (DL) [1]. The main issue associated with Neural Networks and Deep Learning is that they are black boxes. In other words, the results are not known before the NN is trained and evaluated. Hence, experimenting with neural network algorithms in various scenarios is important, as it could produce promising results.

In this particular case, the objective is to automate the process of detecting peaks in a CIR obtained from a 60GHz channel sounder, with 5GHz bandwidth, used for the intra-car channel measurement[2]. The output of the channel measurement is denoted as relative power, representing the ratio of the reflected signal power to the noise, expressed in decibels.

To conduct a comprehensive evaluation of the methods and test their applicability in various measurements an 80GHz channel sounder, which operates within a bandwidth of 1GHz and captures data in the time domain, is also employed. The measurement scenario in this case involved monitoring reflections from a frozen field. The reflections from objects with amplitudes above the noise background can be considered as anomalies. The insights can subsequently be applied in other channel measurement applications such as detection [3].

This paper explores the potential of contemporary Deep Learning techniques for anomaly detection. To achieve this objective, FCNN and LSTM neural networks are employed. The paper is structured in two main sections. The first section outlines the neural network architecture for peak detection, while the second section compares the neural network results with results achieved using filters such as Butterworth, Savitzky-Golay, Bessel/Thomson, and the median filter.

## II. DETECTION OF ANOMALIES USING NEURAL NETWORKS

### A. Basic principles

The core concept of utilizing NNs for anomaly detection is based on training the network to learn the underlying pattern or curve without relying on mathematical models. In this particular case, an autoencoder is used. An autoencoder is a special type of neural network that is trained to copy its input to its output. For example, given an image of a handwritten digit, an autoencoder first encodes the image into a lower dimensional latent representation, then decodes the latent representation back to an image. An autoencoder learns to compress the data while minimizing the reconstruction error [4]. Through the use of the autoencoder, it becomes possible to mitigate the effects of a noise environment. Moreover, the autoencoder does not respond to sudden, rapid changes; rather, it captures the overall trend of the entire function.

This research was funded in part by the National Science Center (NCN), Poland, grant no. 2021/43/I/ST7/03294 (MubaMilWave). For this purpose of Open Access, the author has applied a CC-BY public copyright license to any Author Accepted Manuscript (AAM) version arising from this submission.

## B. Long short-term memory

Long short-term memory networks are a special kind of RNN that can learn long-term dependencies [5]. They find application in various domains, such as predicting time series in medical contexts, as seen in ECG signal analysis [6], [7], forecasting stock market trends [8], or interpreting data outputs from sensors [9].

## C. Fully connected Neural network

Additionally, FCNN has been used to compare its results with the LSTM neural network. A similar FCNN has been previously used in an intra-vehicle scenario [10]. Unlike the LSTM, the FCNN does not consider prior values, which may lead to slightly less accurate outputs. But this attribute paradoxically contributes to anomaly detection, because the anomalies are determined by subtracting the predicted values from the input values.

## III. MEASUREMENT

The channel impulse response from the channel sounder consists of IQ samples [2]. The data is initially transformed into the time domain, which then serves as the input dataset. To prevent the gradient explosion while training the NN, data are scaled with a standard scaler, described in a formula:

$$X_{standardized} = \frac{X - mean(X)}{std(X)} \quad (1)$$

where $X_{standardized}$ is the standardized version of the feature X, $X$ is the original feature, $mean(X)$ is the mean of the feature X, and $std(X)$ is the standard deviation of the feature X.

In this case, a supervised learning method is employed, meaning that the samples and targets are known. The data is then divided into segments, and a sliding window is used to create the sample dataset and target dataset. The method is described with the formula below:

$$input = [X_i, X_{i+1}, X_{i+2}, \ldots X_{i+k}]$$
$$target = X_{i+k+1} \quad (2)$$

where $i$ is the position of sliding windows and $k$ is the number of input samples.

To provide a more accurate description of this process, Figure 1 illustrates an example of the sliding window with 4 input samples.

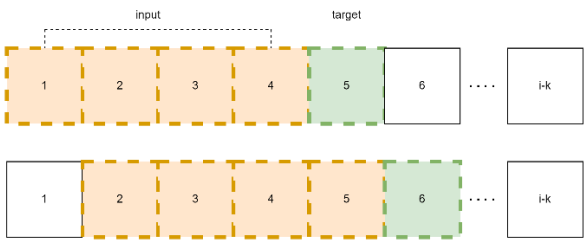

Fig. 1. Example of sliding window with k = 4

The input dataset is afterwards divided into testing sets, each with different sequence lengths of 2, 20, 50, and 100 samples.

## A. Input sequence length evaluation

To analyse the impact of the input sequence length on channel modelling, neural networks with architectures shown in TABLE I and TABLE II are used. The FCNN architecture consists of dense layers, rectified linear activation functions, and dropout functions, which are implemented to prevent overfitting by deactivating a percentage of neurons during training. To properly assess LSTM, all parameters are kept identical to those in FCNN, except for the substitution of dense layers with LSTM layers. Both neural networks share an autoencoder architecture, featuring higher-dimensional layers at the start and end, with a bottleneck in the middle.

TABLE I. FCNN

| Neural network architecture | dense 64, relu, dropout 0.2<br>dense 32, relu, dropout 0.2<br>dense 16, relu,<br>dense 32, relu, dropout 0.2<br>dense 64, relu, dropout 0.2<br>time distributed dense 1 |
|---|---|

TABLE II. LSTM

| Neural network architecture | lstm 64, relu, dropout 0.2<br>lstm 32, relu, dropout 0.2<br>flatten<br>repeat vector 16<br>lstm 32, relu, dropout 0.2<br>lstm 64, relu, dropout 0.2<br>time distributed dense 1 |
|---|---|

The primary distinction between LSTM and FCNN is that FCNN lacks memory cells. Notably, the FCNN yielded good results only with the shortest input sequences, failing to produce satisfactory outcomes with longer input sequences as shown in Figure 2.

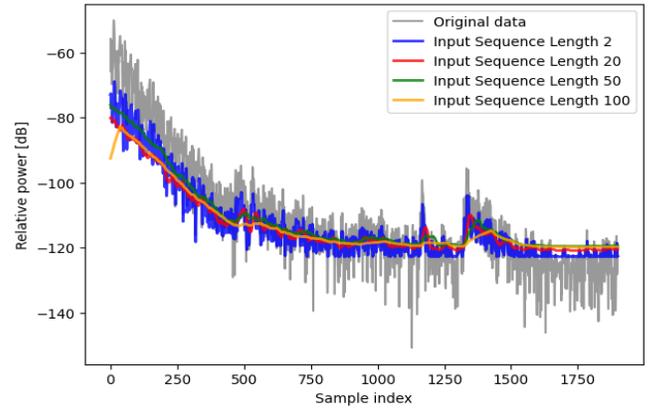

FCNN with varying sequence length

In contrast, the LSTM NN consistently delivered good results with all input lengths as shown in Figure 3. For shorter input sequences, the predicted series closely matches the input sequence. With longer input sequences, the results tend to replicate only the trend of the function, which is desirable for identifying anomalies in the series.

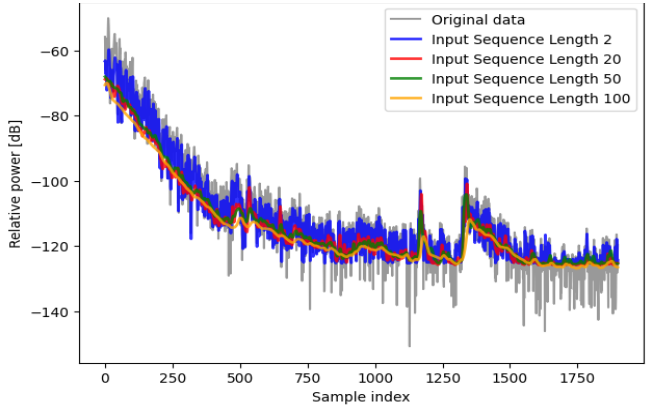

Fig. 2. LSTM with varying sequence length

To conclude the results, the LSTM NN achieved consistent results and therefore it will be used for further analysis of choosing the correct network architecture.

*B. LSTM architecture*

For the upcoming tests, LSTM NN with input sequence of 100 samples is utilized. The sequence of 100 samples yielded the best results in predicting trends rather than replicating the function. The NN is initially trained with architecture in TABLE III and then for each measurement, all of the layers are halved to explore how the network's robustness impacts its performance. Smaller neural networks often exhibit good data generalization capabilities but may struggle to represent the data accurately. Conversely, robust neural networks can effectively learn underlying patterns, although this can sometimes result in overfitting.

TABLE III.   LSTM

| Neural network architecture | lstm 64, relu, dropout 0.2<br>lstm 32, relu, dropout 0.2<br>flatten<br>repeat vector 16<br>lstm 32, relu, dropout 0.2<br>lstm 64, relu, dropout 0.2<br>time distributed dense 1 |
|---|---|

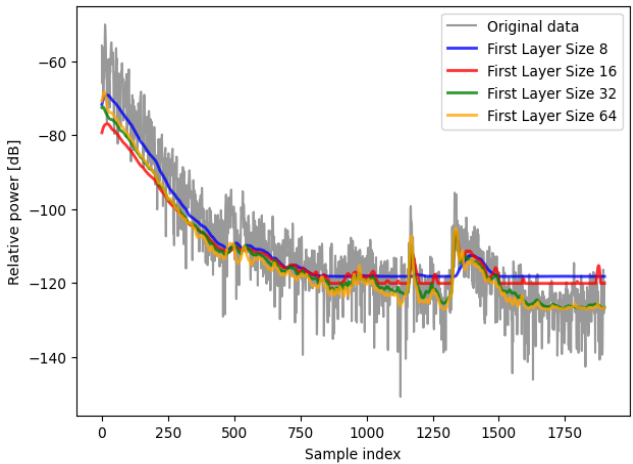

Fig. 3. LSTM of 60GHz channel sounder, with varying NN architecture

The results from Figure 4 indicate that when the size of the neural network decreases, the output values follow the trend of the function rather than replicating the function itself. To further observe the behavior of the NN architecture, a second test was conducted using an 80GHz channel sounder system for measuring a frozen field scenario. The results shown in Figure 5 demonstrate that chosen LSTM architercure is convenient for processing data from different CIR measurements

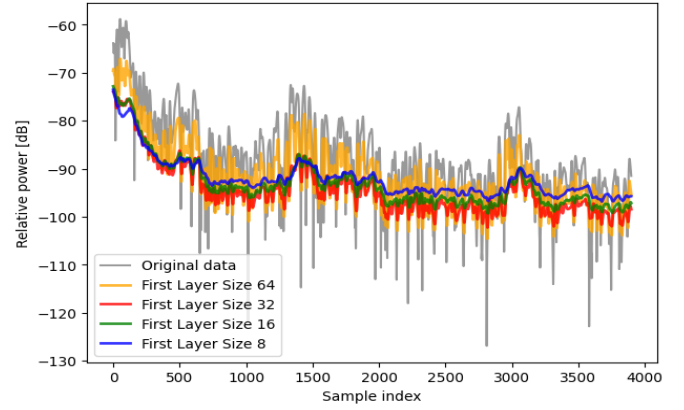

Fig. 4. LSTM of 80GHz channel sounder, with varying NN architecture

The drawback of using neural networks is that they need to be trained for each signal impulse response. The training process consisted of 20 epochs, with each epoch requiring 2 seconds for training.

## IV. ANOMALY DETECTION

*A. Input-Output analysis*

To detect anomalies, we calculate the distance between each two points in a graph. Using mean square error is not suitable for this application because it includes negative peaks. The solution for this calculation is a simple subtraction of the predicted function from the input function.

$$Error = input(x) - predicted(x) \quad (3)$$

The graph of the distances between the points is shown in Figure 6. In this picture, the peaks are clearly visible. To fully automate the process of identifying peaks or even entire clusters, a thresholding method needs to be applied.

*B. DBSCAN*

To separate anomalies from the noise a clustering method Density-Based Spatial Clustering of Applications with Noise (DBSCAN) from scikit learn library [11] with default parameters will be used. DBSCAN algorithm defines clusters as continuous regions of high density. For each instance, the algorithm counts how many instances are located within a small distance from it [1]. In the following figures, the graphs of the original and predicted data are displayed. Additionally, there is a figure showing the calculated distances between the functions. The distinguishing colors in the graph below represent anomalies obtained from the default settings of the DBSCAN function, with the min_samples value set to two.

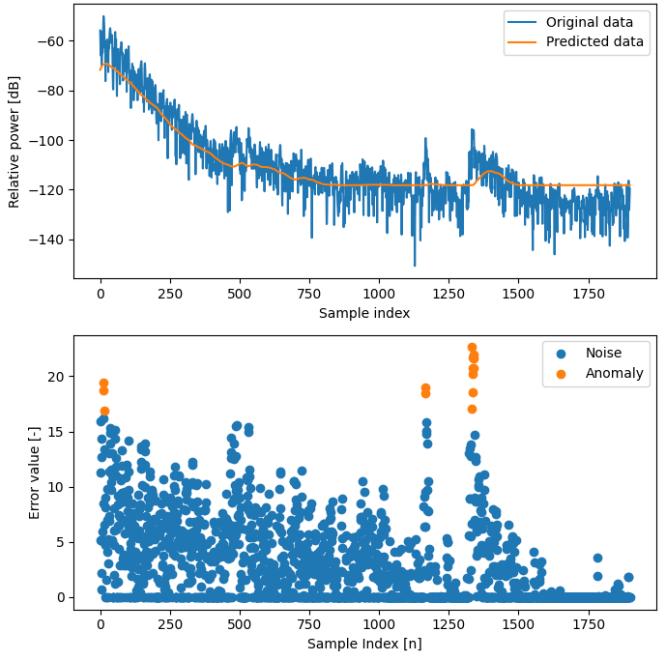

Fig. 5. CIR 1 - Detected anomalies after applying NN and DBSCAN

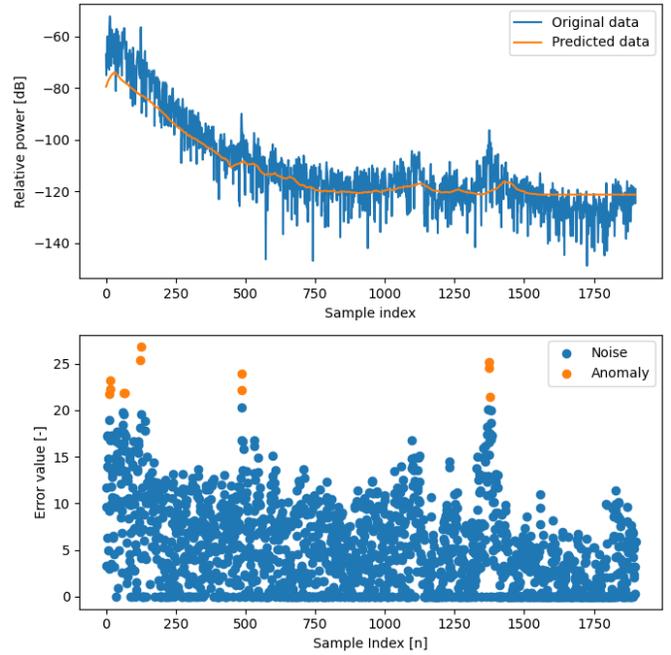

Fig. 7. CIR 3 - Detected anomalies after applying NN and DBSCAN

The results using LSTM NN and DBSCAN show that this method is effective for detecting peaks and assigning clusters in a noise environment, however, this method experiences notable errors within the first 100 samples of the graph as shown in Figure 6 and Figure 8. These discrepancies might be caused by the initial lack of context at the beginning of the graph prediction.

## V. COMPARISON TO SIGNAL FILTERS

To further validate its effectiveness, the results from LSTM NN presented in TABLE IV are compared to commonly used filters from the SciPy library [12]. In this case, Butterworth, Savitzky-Golay, Bessel/Thomson, and median filters were verified.

TABLE IV. LSTM ARCHITECTURE

| | |
|---|---|
| Neural network architecture | lstm 8, relu, dropout 0.2 |
| | lstm 4, relu, dropout 0.2 |
| | flatten |
| | repeat vector 2 |
| | lstm 4, relu, dropout 0.2 |
| | lstm 8, relu, dropout 0.2 |
| | time distributed dense 1 |

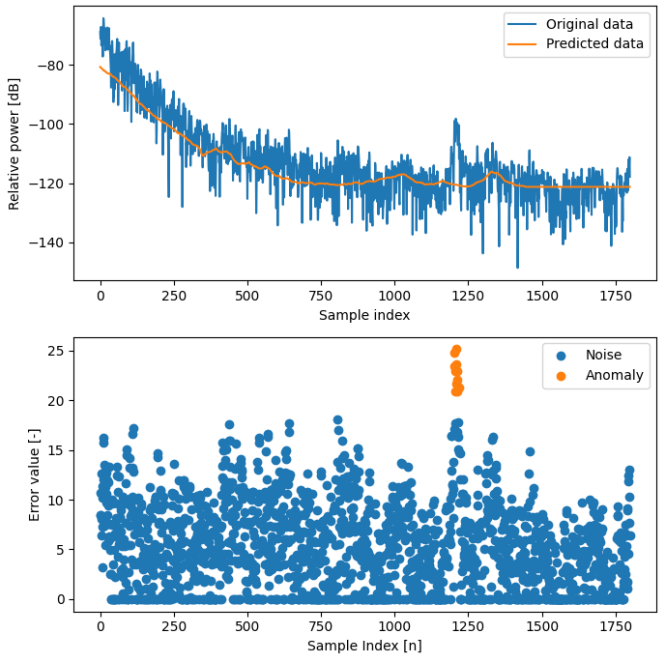

Fig. 6. CIR 2 - Detected anomalies after applying NN and DBSCAN

To achieve desirable results, it is necessary to correctly configure filter parameters. As there is no automated method for parameter configuration, they were chosen manually, beginning with conservative values, and subsequently fine-tuned to visually obtain the best results. This manual selection of parameters contrasts with the objective of automation. The selected parameters are listed in TABLE V. The Following figures illustrate the impact of various filters on peak cluster detection.

TABLE V. FILTER PARAMETERS

| Savitzky-Golay | *Window length = 100, Polynomial order = 5* |
| Butterworth | *Order of the filter = 10, Cutoff frequency = 0.04*Fs, Filter type = Lowpass* |
| Bessel/Thomson | *Order of the filter = 4, Cutoff frequency = 0.1*Fs, Filter type = Lowpass* |
| Median filter | *Input array length = 100* |

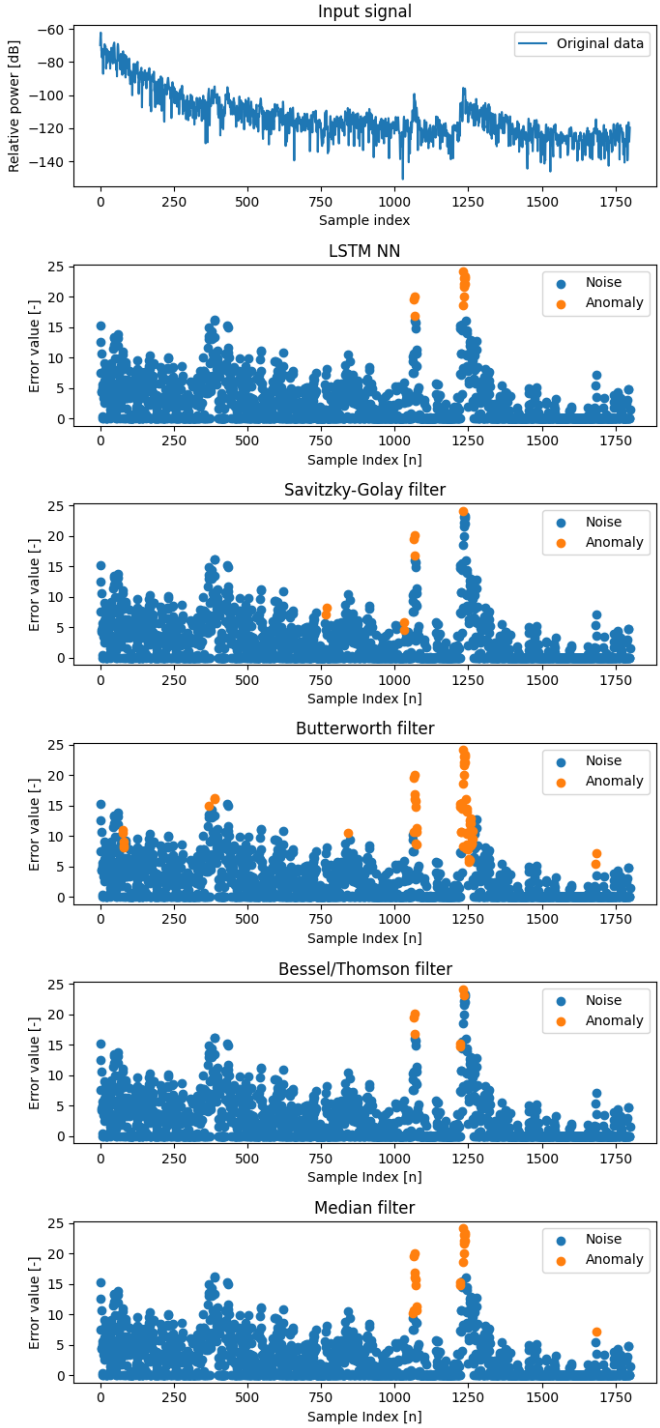

Fig. 8. CIR 1 - Comparison of LSTM NN and other filters

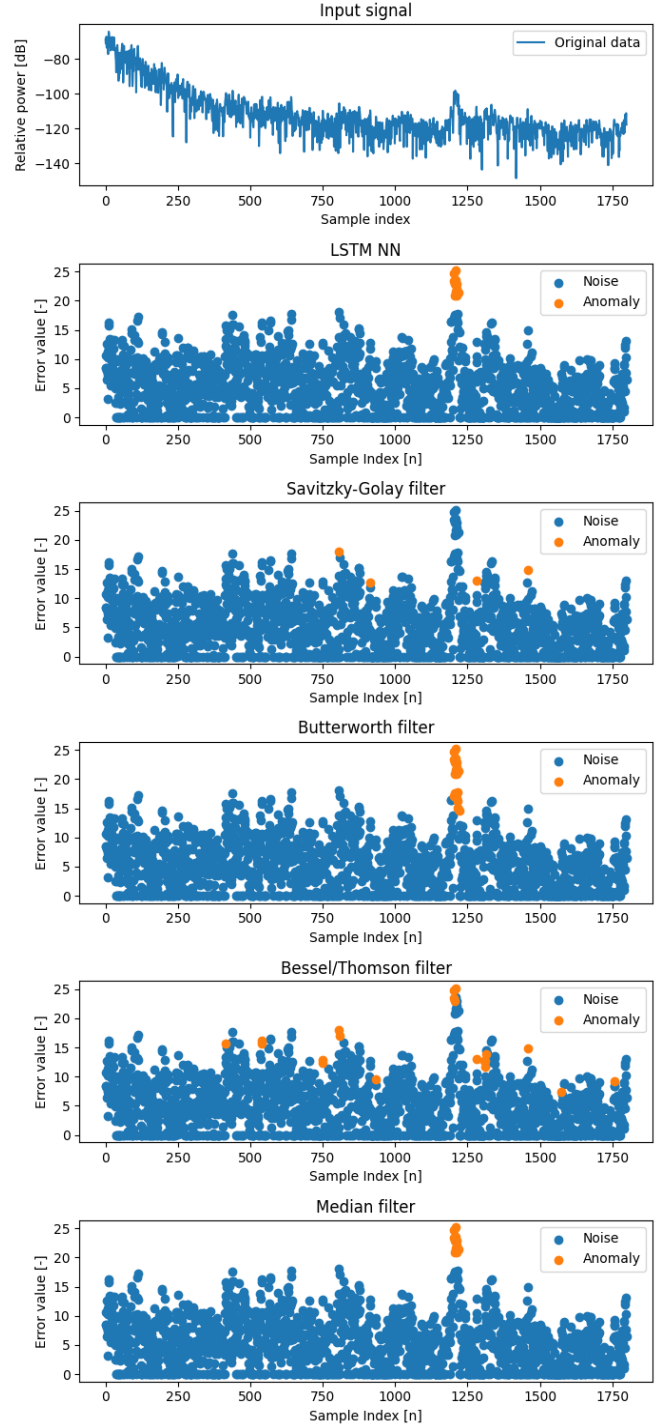

Fig. 9. CIR 2 - Comparison of LSTM NN and other filters

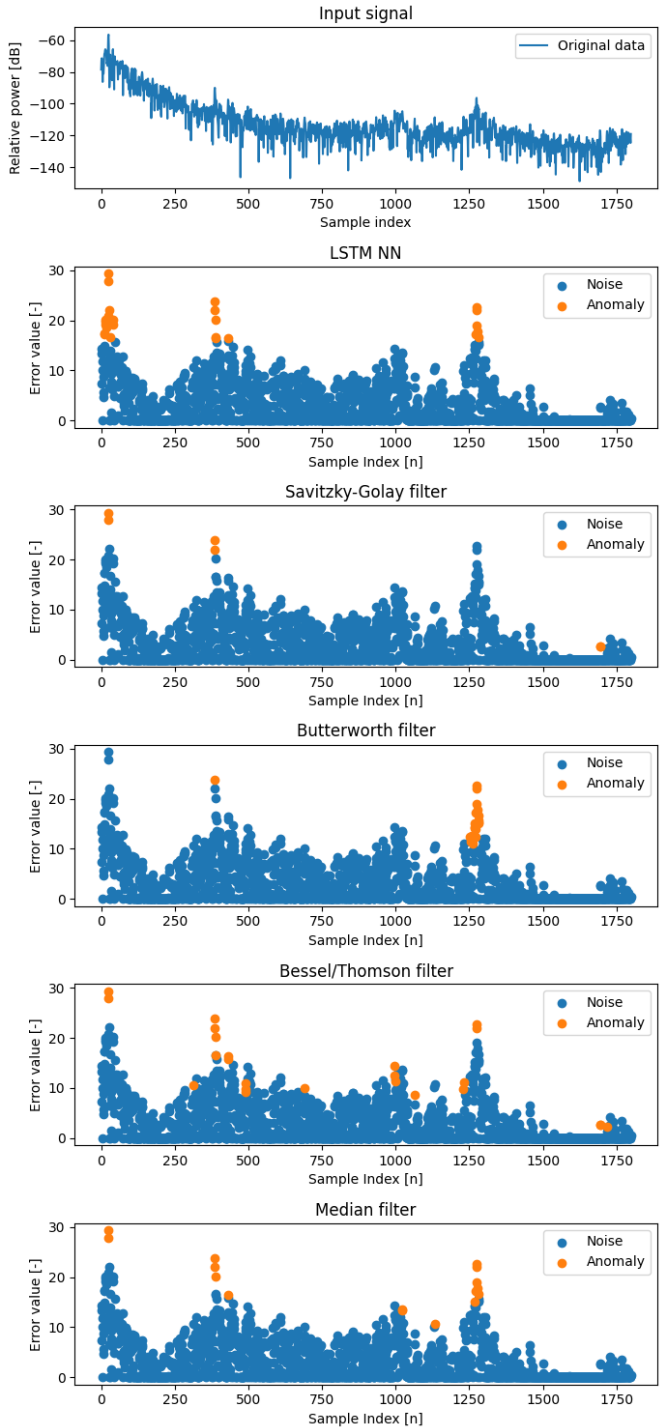

Fig. 10. CIR 3 - Comparison of LSTM NN and other filters

The results show that the highest precision is achieved when using DBSCAN with LSTM NN and Median filter. Although some filters produce favourable outcomes, they are inconsistent in their results. It is also notable, that LSTM NN results vary from the previous measurement, despite the same neural network architecture. This is due to the fact that NN can achieve different results with each training cycle These findings indicate that while neural networks may be suitable for this application, further testing is required. The primary area for improvement lies in integrating more input data or exploring combinations of digital filters and neural networks.

## VI. Acknowledgement

The research was conducted within the framework of a research grant project funded by the Czech Science Foundation, under Lead Agency Project No. 22-04304L, Mutli-band prediction of millimetre-wave propagation effects for dynamic and fixed scenarios in rugged time-varying environments.

## VII. CONCLUSION

This study deals with the application of Deep Learning techniques, specifically Long Short-Term Memory (LSTM) and Fully Connected Neural Networks (FCNN), for the purpose of anomaly detection in millimeter wave channel impulse response. The core principle involves training neural networks to learn trends of the reflected signals impulse response and to automate the process of peak detection.

The research reveals that LSTM neural networks outperformed FCNN in scenarios of predicting signal impulse response. The LSTM demonstrated robustness in predicting trends. In contrast, FCNN exhibited limitations with longer input sequences, leading to less accurate results.

The study also employs the DBSCAN clustering method to further distinguish anomalies from noise, showing that the combination of LSTM and DBSCAN is effective in detecting peaks in noise environments.

Furthermore, the results given by the LSTM neural network and DBSCAN are compared with traditional signal filters like Butterworth, Savitzky-Golay, Bessel/Thomson, and median filters. The deployment of these filters comes with a manual selection of parameters such as cutoff frequency, filter order, and window length, which contradicts the goal of automation. Based on the visual comparison, the most effective methods for peak detection are either median filtering or the use of LSTM neural networks. However, it's important to note that a drawback of neural networks is the necessity to train them for each impulse response, which consumes approximately 40 seconds. Additionally, due to the training process, there might be slight variations in the results with each new training session.

The further scope of research could divide into two groups. firstly, exploring the application of the latest neural network architectures or creating ensemble models that combine neural networks with machine learning models to achieve improved output trend representations and secondly, examining novel methods for evaluating the anomalies from the CIR input and its calculated trend function.